\documentclass [11pt]{article}
\title{"Pseudo" pseudo forces in gauge synchronization theories}
\author{Robert D. Klauber
\\1100 University Manor Dr., 38B, Fairfield, Iowa 52556
\\klauber@.iowatelecom.net, permanent: rklauber@.netscape.net
\\(Omit dot after @ in above, used to thwart bots seeking spam addresses)}

\date{March 29, 2004}

\usepackage{graphicx}
\usepackage{longtable}

\begin{document}

\maketitle

\begin{abstract}

The role of extra terms in Newton's second law that arise as a result of 
non-Einstein synchronization is explored. Although such extra terms have 
been interpreted as pseudo forces that constrain physical theory to a unique 
method of synchronization (the Einstein, or standard, synchronization), a 
fully relativistic analysis indicates that such extra terms do not 
invalidate the conventionalist thesis of synchronization.

\end{abstract}

\section{Introduction}
In a recent article Ohanian\cite{Ohanian:2002} contends that 
synchronizations other than the standard Einstein synchronization (E 
synchronization) result in pseudo force terms arising in Newton's second 
law, and that because pseudo forces are not observed in inertial frames, 
non-standard synchronizations of the Reichenbach\cite{Reichenbach:1927} 
type (R sychronizations) can not represent the true state of inertial frames 
in the physical world. That is, synchronization is not simply a physically 
meaningless gauge to be applied to clock settings, but is constrained by 
nature to be unique for inertial frames.

As a counterpoint, one could argue that if clocks in various places have 
different settings, and~one is measuring~the motion of a body on which a 
force is applied, the body~will seem to take a different amount of time to 
get from~point A to point B, depending on the settings on the different sets 
of clocks.~ One measures acceleration with clocks and rods, and having 
different sets of clocks means having different numerical measurements of 
acceleration \textbf{a}, for the same particle motion. So 
\textbf{F}=m\textbf{a} would not be expected to describe the motion~caused 
by~a force~\textbf{F} for any but the standard synchronization,~and we would 
need additional term(s)/factor(s) to correct for the difference between 
\textbf{F} and m\textbf{a} that arises.

If we simply transformed the standard 4D~relativistic form of the 2nd law to 
new synchronization coordinates, and took the low speed limit, we would get 
the appropriate non-standard coordinate form of the 2nd law, which would 
have extra term(s)/factor(s) that could be interpreted as pseudo force(s).~ 
We would then have a whole (different) set of dynamical laws that would be 
consistent internally and in agreement with experiment (provided the same 
synchronization were used on the experimental clocks.) So does this, or does 
this not, mean that R synchronizations are invalid? Does the presence of 
apparent pseudo forces in the equation of motion doom the conventionalist 
interpretation of synchronization? This article addresses these questions, 
and concludes that the conventionalist thesis is not only still alive, but 
also still kicking.

In Section \ref{sec:psuedo}, as a reference example of pseudo 
forces, Coriolis and centrifugal forces in rotation are derived using 
relativity theory. The same procedure is then applied in Section 
\ref{sec:pseudo} to coordinate frames with R synchronization. The 
results are compared in Section \ref{sec:alternative} with Ohanian's 
derivation and in Section \ref{sec:inertial} with various definitions 
for ``inertial frame''. Philosophical implications are addressed in Section 
\ref{sec:philosophy}, and conclusions drawn in Section 
\ref{sec:conclusion}.

\section{Psuedo Forces Example: Rotation}
\label{sec:psuedo}
Pseudo forces in accelerating frames arise (mathematically) from the 
non-Lorentzian metric. The following is well known and presented for 
background, as well as for later comparison to the non-standard 
synchronization case.

For rotation, with familiar symbols for cylindrical coordinates and the 
coordinate transformation
\begin{equation}
\label{eq1}
\begin{array}{l}
 cT=ct\quad \quad \quad \quad \quad \mbox{(a)} \\ 
 R=r\quad \quad \quad \quad \;\;\;\quad \mbox{(b)} \\ 
 \Phi =\phi +\omega t\quad \quad \quad \;\;\mbox{(c)} \\ 
 Z=z\quad \quad \quad \quad \quad \;\;\;\mbox{(d)} \\ 
 \end{array},
\end{equation}
where upper case refer to the lab frame, and lower case to the rotating 
frame, the metric and its inverse\cite{Found:1} are
\begin{equation}
\label{eq2}
g_{\alpha \beta } =\left[ {{\begin{array}{*{20}c}
 {-(1-\textstyle{{r^2\omega ^2} \over {c^2}})} \hfill & 0 \hfill & 
{\textstyle{{r^2\omega } \over c}} \hfill & 0 \hfill \\
 0 \hfill & 1 \hfill & 0 \hfill & 0 \hfill \\
 {\textstyle{{r^2\omega } \over c}} \hfill & 0 \hfill & {r^2} \hfill & 0 
\hfill \\
 0 \hfill & 0 \hfill & 0 \hfill & 1 \hfill \\
\end{array} }} \right]\quad g^{\alpha \beta }=\left[ {{\begin{array}{*{20}c}
 {-1} \hfill & 0 \hfill & {\textstyle{\omega \over c}} \hfill & 0 \hfill \\
 0 \hfill & 1 \hfill & 0 \hfill & 0 \hfill \\
 {\textstyle{\omega \over c}} \hfill & 0 \hfill & {(1-\textstyle{{r^2\omega 
^2} \over {c^2}})/r^2} \hfill & 0 \hfill \\
 0 \hfill & 0 \hfill & 0 \hfill & 1 \hfill \\
\end{array} }} \right].
\end{equation}
The only non-zero Christoffel symbols, found from
\begin{equation}
\label{eq3}
\Gamma _{\alpha \beta g} =\textstyle{1 \over 2}\left( {g_{\alpha \beta 
,\gamma } +g_{\alpha \gamma ,\beta } -g_{\beta \gamma ,\alpha } } 
\right);\,\,\,\,\,\,\Gamma ^\alpha _{\beta g} =g^{\alpha \mu }\Gamma _{\mu 
\beta g} \,,
\end{equation}
are
\begin{equation}
\label{eq4}
\Gamma ^r_{tt} =\Gamma ^1_{00} =-\frac{\omega 
^2r}{c^2};\,\,\,\,\,\,\,\,\Gamma ^\theta _{tr} =\Gamma ^2_{01} =\frac{\omega 
}{cr};\,\,\,\,\,\,\,\,\,\Gamma ^r_{t\theta } =\Gamma ^1_{02} =-\frac{\omega 
r}{c}.
\end{equation}
The equation of motion for a geodesic particle is
\begin{equation}
\label{eq5}
\frac{d^2x^\alpha }{d\tau ^2}+\Gamma ^\alpha _{\beta \gamma } \frac{dx^\beta 
}{d\tau }\frac{dx^\gamma }{d\tau }=0.
\end{equation}
The relevant 4-velocities are
\begin{equation}
\label{eq6}
\begin{array}{l}
 u^0=\frac{dx^0}{d\tau }=\frac{cdt}{d\tau }=\frac{c}{\sqrt {1-\omega 
^2r^2/c^2} } \\ 
 u^1=\frac{dx^1}{d\tau }=\frac{dr}{d\tau } \\ 
 u^2=\frac{dx^2}{d\tau }=\frac{d\theta }{d\tau }\,\,. \\ 
 \end{array}
\end{equation}
\subsection{Radial Direction}
\label{subsec:radial}
For the $x^{1}=r$ direction, the equation of motion (\ref{eq5}) becomes
\begin{equation}
\label{eq7}
a^r=a^1=\frac{\omega ^2r}{\left( {1-\omega ^2r^2/c^2} \right)}+\frac{\omega 
r}{\sqrt {1-\omega ^2r^2/c^2} }\frac{d\theta }{d\tau }=\underbrace 
{\frac{\omega ^2r}{\left( {1-\omega ^2r^2/c^2} 
\right)}}_{Centrifugal\,\,accel}+\underbrace {\frac{\omega u^{\hat {\theta 
}}}{\sqrt {1-\omega ^2r^2/c^2} }}_{Coriolis\,\,accel}
\end{equation}
where $u^{\hat {\theta }}=\sqrt {g_{\theta \theta } } d\theta /d\tau 
=rd\theta /d\tau $ is the physical velocity (i.e., measured in m/s using 
standard meter sticks) of the particle in the \textit{$\theta $} direction relative to the 
rotating frame. Since the particle is undergoing geodesic motion, as seen 
from the rotating frame there is acceleration relative to the rotating frame 
coordinates. For a particle fixed at constant radius $r$ in the rotating frame, 
centrifugal and Coriolis pseudo forces equal to the mass times the terms on 
the RH side of (\ref{eq7}) would appear to arise.

\subsection{Tangential Direction}
\label{subsec:tangential}
For the $x^{2}$ = \textit{$\theta $} direction, the equation of motion (\ref{eq5}) becomes
\begin{equation}
\label{eq8}
a^\theta =a^2=\frac{d^2\theta }{d\tau ^2}=-\frac{\omega }{r\sqrt {1-\omega 
^2r^2/c^2} }\frac{dr}{d\tau }=-\frac{\omega }{r\sqrt {1-\omega ^2r^2/c^2} 
}u^{\hat {r}}
\end{equation}
where $u^{\hat {r}}=\sqrt {g_{rr} } dr/d\tau =dr/d\tau $ is the physical 
velocity in the radial direction relative to the rotating frame.

The physical (measured in m/s$^{2})$ value for the tangential acceleration 
is
\begin{equation}
\label{eq9}
a^{\hat {\theta }}=\sqrt {g_{\theta \theta } } a^\theta =ra^\theta 
=\underbrace {-\frac{\omega }{\sqrt {1-\omega ^2r^2/c^2} }u^{\hat 
{r}}}_{Coriolis\,\,accel}.
\end{equation}
\section{Pseudo Forces and Non-standard Synchronization Gauges}
\label{sec:pseudo}
\subsection{Relativity, Geodesics, and Pseudo Accelerations}
From eqs (\ref{eq5}) or (\ref{eq6}) of Anderson et al\cite{Anderson:1998}, the 
transformation from Lorentz coordinates (with standard, Einstein 
synchronization) to non-standard R synchronization coordinates is
\begin{equation}
\label{eq10}
\begin{array}{l}
 c\tilde {t}=ct-kx\quad \quad \quad \quad \;(a) \\ 
 \tilde {x}=x\quad \quad \quad \quad \quad \quad \quad (b) \\ 
 \tilde {y}=y\quad \quad \quad \quad \quad \quad \quad (c) \\ 
 \tilde {z}=z\quad \quad \quad \quad \quad \quad \quad (d), \\ 
 \end{array}
\end{equation}
where $k$ is a constant and $-1<k<1$. One should note that where we and 
Anderson et al employ $k$, Ohanian uses \textit{ck}. From (\ref{eq10}), one can 
derive\cite{Ref:1} the metric for the R synchronization and its inverse,
\begin{equation}
\label{eq11}
\tilde {g}_{\alpha \beta } =\left[ {{\begin{array}{*{20}c}
 {-1} \hfill & {-k} \hfill & 0 \hfill & 0 \hfill \\
 {-k} \hfill & {1-k^2} \hfill & 0 \hfill & 0 \hfill \\
 0 \hfill & 0 \hfill & 1 \hfill & 0 \hfill \\
 0 \hfill & 0 \hfill & 0 \hfill & 1 \hfill \\
\end{array} }} \right]\quad 
\,\,\,\,\,\,\,\,\,\,\,\,\,\,\,\,\,\,\,\,\,\,\,\,\,\tilde {g}^{\alpha \beta 
}=\left[ {{\begin{array}{*{20}c}
 {-(1-k^2)} \hfill & {-k} \hfill & 0 \hfill & 0 \hfill \\
 {-k} \hfill & 1 \hfill & 0 \hfill & 0 \hfill \\
 0 \hfill & 0 \hfill & 1 \hfill & 0 \hfill \\
 0 \hfill & 0 \hfill & 0 \hfill & 1 \hfill \\
\end{array} }} \right]\quad .
\end{equation}
Repeating the procedure used in rotation to find pseudo forces, one finds, 
because the metric is constant in spacetime, that all Christoffel symbols 
(\ref{eq3}) are zero, i.e.,
\begin{equation}
\label{eq12}
\Gamma ^\alpha _{\beta \gamma } 
=0\,\,\,\,\,\,\,\,\mbox{for}\,\mbox{all}\,\,\alpha \mbox{,}\beta 
\mbox{,}\gamma .
\end{equation}
Thus, from (\ref{eq5}) we must have
\begin{equation}
\label{eq13}
\frac{d^2\tilde {x}^\alpha }{d\tau 
^2}=0\,\,\,\,\,\,\mbox{for}\,\mbox{all}\,\alpha ,
\end{equation}
there are no pseudo accelerations, and there can be no pseudo forces, in 
apparent contradiction to Ohanian's claim.

\subsection{Proper Time, Applied and Pseudo Forces}
\label{subsec:proper}
However, to be precise, Ohanian claimed only that motion of a non-goedesic 
(accelerated as seen in a Lorentz coordinate frame by a force \textbf{F}) 
would be modified by additional pseudo forces (as seen in the R 
synchronization coordinate frame.) Thus, consider the more general form of 
the 4D equation (\ref{eq5}) when forces are present,
\begin{equation}
\label{eq14}
m\left( {\frac{d^2x^\alpha }{d\tau ^2}+\Gamma ^\alpha _{\beta \gamma } 
\frac{dx^\beta }{d\tau }\frac{dx^\gamma }{d\tau }} \right)=F^\alpha ,
\end{equation}
where $F^{\alpha }$ is the 4D generalized covariant force. For the Lorentz 
coordinate frame, the Christoffel symbols in (\ref{eq14}) are zero, and we have
\begin{equation}
\label{eq15}
ma_\tau ^\alpha =m\frac{d^2x^\alpha }{d\tau ^2}=F^\alpha .
\end{equation}
Transform (\ref{eq15}) to the R synchronization coordinates via the matrix inherent 
in (\ref{eq10}), 
\begin{equation}
\label{eq16}
m\left[ {{\begin{array}{*{20}c}
 {\tilde {a}_\tau ^0} \hfill \\
 {\tilde {a}_\tau ^1} \hfill \\
 {\tilde {a}_\tau ^2} \hfill \\
 {\tilde {a}_\tau ^3} \hfill \\
\end{array} }} \right]=m\left[ {{\begin{array}{*{20}c}
 1 \hfill & {-k} \hfill & 0 \hfill & 0 \hfill \\
 0 \hfill & 1 \hfill & 0 \hfill & 0 \hfill \\
 0 \hfill & 0 \hfill & 1 \hfill & 0 \hfill \\
 0 \hfill & 0 \hfill & 0 \hfill & 1 \hfill \\
\end{array} }} \right]\left[ {{\begin{array}{*{20}c}
 {a_\tau ^0} \hfill \\
 {a_\tau ^1} \hfill \\
 {a_\tau ^2} \hfill \\
 {a_\tau ^3} \hfill \\
\end{array} }} \right]=\left[ {{\begin{array}{*{20}c}
 1 \hfill & {-k} \hfill & 0 \hfill & 0 \hfill \\
 0 \hfill & 1 \hfill & 0 \hfill & 0 \hfill \\
 0 \hfill & 0 \hfill & 1 \hfill & 0 \hfill \\
 0 \hfill & 0 \hfill & 0 \hfill & 1 \hfill \\
\end{array} }} \right]\left[ {{\begin{array}{*{20}c}
 {F^0} \hfill \\
 {F^1} \hfill \\
 {F^2} \hfill \\
 {F^3} \hfill \\
\end{array} }} \right]=\left[ {{\begin{array}{*{20}c}
 {\tilde {F}^0} \hfill \\
 {\tilde {F}^1} \hfill \\
 {\tilde {F}^2} \hfill \\
 {\tilde {F}^3} \hfill \\
\end{array} }} \right].
\end{equation}
Thus, in 3D, nothing is really changed (the same is true for the 0$^{th}$ 
component as the $k$ terms drop out), and we have
\begin{equation}
\label{eq17}
m\tilde {a}_\tau ^i=m\frac{d^2\tilde {x}^i}{d\tau ^2}=\tilde 
{F}^i=F^i\,\,\,\,\,\,\,\,\,\,\,\,\,(i=1,2,3),
\end{equation}
i.e., there are no new pseudo forces arising in the R synchronization 
coordinate system.

\subsection{Coordinate Time, Applied and Pseudo Forces}
\label{subsec:coordinate}
Note, however, that the time variable in (\ref{eq17}) is the proper time on the 
particle, which, of course, is invariant, i.e., the same in any coordinate 
system with any time coordinate synchronization. One could (and in the 
present context, \textit{should}) ask what the equation of motion for a geodesic would be 
in terms of the coordinate time $\tilde {t}$ of the R synchronization of 
(\ref{eq10}).

From (\ref{eq11})
\begin{equation}
\label{eq18}
c^2d\tau ^2=-\tilde {g}_{\alpha \beta } d\tilde {x}^\alpha d\tilde {x}^\beta 
=c^2d\tilde {t}^2+2ckd\tilde {x}^1d\tilde {t}-(1-k^2)\left( {d\tilde {x}^1} 
\right)^2-\left( {d\tilde {x}^2} \right)^2-\left( {d\tilde {x}^3} 
\right)^2.
\end{equation}
Dividing (\ref{eq18}) by $d\tilde {t}$ and solving for \textit{d$\tau $} one finds
\begin{equation}
\label{eq19}
d\tau =\sqrt {\left( {1+\frac{k}{c}\tilde {v}^1} \right)^2-\frac{\tilde 
{v}^2}{c^2}} \,d\tilde {t}\cong \,\left( {1+\frac{k}{c}\tilde {v}^1} 
\right)d\tilde {t}
\end{equation}
with the approximation on the RH to first order and
\begin{equation}
\label{eq20}
\tilde {v}^1=\frac{d\tilde {x}^1}{d\tilde {t}}\,\,\,\,\,\,\,\,\,\,\,\tilde 
{v}^2=\left( {\frac{d\tilde {x}^1}{d\tilde {t}}\,} \right)^2+\left( 
{\frac{d\tilde {x}^2}{d\tilde {t}}\,} \right)^2+\left( {\frac{d\tilde 
{x}^3}{d\tilde {t}}\,} \right)^2.
\end{equation}
In what follows, one could use the exact expression of (\ref{eq19}) and take the 
first order limit of the final result, though for simplicity we shall from 
the beginning simply use the approximation on the RH of (\ref{eq19}). Noting that
\begin{equation}
\label{eq21}
\tilde {u}^i=\frac{d\tilde {x}^i}{d\tau }\cong \frac{1}{\left( 
{1+\frac{k}{c}\tilde {v}^1} \right)\,}\left( {\frac{d\tilde {x}^i}{d\tilde 
{t}}} \right)=\frac{\tilde {v}^i}{\left( {1+\frac{k}{c}\tilde {v}^1} 
\right)\,},
\end{equation}
we can re-express (\ref{eq17}) as
\begin{equation}
\label{eq22}
m\frac{d^2\tilde {x}^i}{d\tau ^2}=F^i=m\frac{d\tilde {u}^i}{d\tau }\cong 
\frac{m}{\left( {1+\frac{k}{c}\tilde {v}^1} \right)\,}\frac{d}{d\tilde 
{t}}\left( {\frac{\tilde {v}^i}{\left( {1+\frac{k}{c}\tilde {v}^1} 
\right)\,}} \right).
\end{equation}
Carrying out the derivative above, one gets
\begin{equation}
\label{eq23}
F^i\cong \frac{m\tilde {a}^i}{\left( {1+\frac{k}{c}\tilde {v}^1} 
\right)^2\,}-\frac{m\tilde {v}^i}{\left( {1+\frac{k}{c}\tilde {v}^1} 
\right)^3\,}\left( {\frac{k\tilde {a}^1}{c}} \right).
\end{equation}
Rearranged and expressed in vector notation, this becomes the same as 
Ohanian's equation (\ref{eq9}), 
\begin{equation}
\label{eq24}
m{\rm {\bf \tilde {a}}}=\left( {1+\frac{{\rm {\bf k}}\cdot {\rm {\bf \tilde 
{v}}}}{c}} \right)^2{\rm {\bf F}}+\frac{\frac{{\rm {\bf k}}\cdot {\rm {\bf 
\tilde {a}}}}{c}}{1+\frac{{\rm {\bf k}}\cdot {\rm {\bf \tilde 
{v}}}}{c}}m{\rm {\bf \tilde {v}}},
\end{equation}
where here and from henceforth we drop the approximate symbol on the equal 
signs, and we remind the reader that Ohanian's notation has \textbf{k}, 
where we have \textbf{k}/$c$ (conforming with Anderson et al). Performing an 
inner product of (\ref{eq24}) with \textbf{k}, solving for ${\rm {\bf k}}\cdot {\rm 
{\bf \tilde {a}}}$, and substituting the result back into (\ref{eq24}), one obtains 
Ohanian's equation (\ref{eq10}),
\begin{equation}
\label{eq25}
m{\rm {\bf \tilde {a}}}=\left( {1+\frac{{\rm {\bf k}}\cdot {\rm {\bf \tilde 
{v}}}}{c}} \right)^2\left( {{\rm {\bf F}}+\frac{{\rm {\bf k}}\cdot {\rm {\bf 
F}}}{c}{\rm {\bf \tilde {v}}}} \right).
\end{equation}
At first blush, (\ref{eq25}) appears to contain extra terms not found in the more 
familiar form of the second law of Newton's dynamics, and one might be 
tempted to interpret these terms as pseudo forces. However, all the 
unfamiliar terms and factors in (\ref{eq25}) are due solely to the relationship 
between proper and coordinate standard times being different in the E and R 
systems. The non-standard expression (\ref{eq19}) for proper time in terms of R 
system clock times lies at the root of (\ref{eq25}). Since proper times are the same 
in both systems, the difference is due to the difference in coordinate 
clocks, i.e., to the system of synchronization chosen. The apparent dynamic 
effect in (\ref{eq25}) is thus not ``real'', being based wholly in the definitions 
chosen for clock settings.

\section{Alternative Synchronization and Newton's Laws}
\label{sec:alternative}
\subsection{Pseudo Forces and Accleration}
To gain a feeling for the physical implications of the relation (\ref{eq25}), we 
break that relation into components of acceleration parallel, and 
perpendicular, to the direction of alternative synchronization (the 
\textbf{k} direction.) 

\subsubsection{Acceleration Parallel to k}
For the parallel direction,
\begin{equation}
\label{eq26}
m\tilde {a}_k =\left( {1+\frac{k\tilde {v}_k }{c}} \right)^2\left( {F_k 
+\frac{kF_k }{c}\tilde {v}_k } \right).
\end{equation}
Consider the case where velocity is in the \textbf{k} direction. The 
equation of motion (\ref{eq26}) then becomes
\begin{equation}
\label{eq27}
m\tilde {a}_k =\left( {1+\frac{k\tilde {v}_k }{c}} \right)^3F_k ,
\end{equation}
and the unusual factor in parentheses is simply a correction to the 
acceleration due to the difference in clock readings as the object moves in 
the \textbf{k} direction. That is, the E synchronization acceleration 
$a_{k}$ takes on the value of $\tilde {a}_k $ in the R synchronization scheme 
as
\begin{equation}
\label{eq28}
a_k \to \tilde {a}_k =\left( {1+\frac{k\tilde {v}_k }{c}} \right)^3\,a_k 
=\left( {1+\frac{k\tilde {v}_k }{c}} \right)^3\,\frac{F_k }{m},
\end{equation}
and no actual change in force, i.e. no pseudo force, exists, either in the 
parallel or transverse direction.

For the case where velocity is solely (and instantaneously) perpendicular to 
\textbf{k}, the equation of motion (\ref{eq26}) becomes
\begin{equation}
\label{eq29}
m\tilde {a}_k =F_k ,
\end{equation}
which is not unusual, and certainly contains no pseudo force. With no 
non-standard clock settings in the transverse direction (the direction of 
the particle velocity), no difference arises from the standard form of 
Newton's second law.

\subsubsection{Acceleration Perpendicular to k}
\label{subsubsec:acceleration}
In component form, the vector equation of motion (\ref{eq25}) for the direction 
perpendicular to the R synchronization is
\begin{equation}
\label{eq30}
m\tilde {a}_\bot =\left( {1+\frac{k\tilde {v}_k }{c}} \right)^2\left( 
{F_\bot +\frac{kF_k }{c}\tilde {v}_\bot } \right).
\end{equation}
For velocity solely in the \textbf{k} direction, this becomes
\begin{equation}
\label{eq31}
m\tilde {a}_\bot =\left( {1+\frac{k\tilde {v}_k }{c}} \right)^2F_\bot 
\end{equation}
for which the bracketed quantity is once again a correction to the 
acceleration due to the particle velocity in the \textbf{k} direction and 
the concomitant change in measured values of acceleration using different 
clocks.

For velocity solely perpendicular to \textbf{k}, (\ref{eq30}) becomes
\begin{equation}
\label{eq32}
m\tilde {a}_\bot =F_\bot +\frac{kF_k }{c}\tilde {v}_\bot ,
\end{equation}
which seems to contain the very unusual (and certainly non-Newtonian) 
characteristic of having force in the \textbf{k} direction $F_{k}$ contribute 
to acceleration in the direction transverse to \textbf{k}. However, one has 
to recognize that the force $F_{k}$ will accelerate the object in the 
\textbf{k} direction via (\ref{eq29}), thereby giving it motion in that direction. 
So, when subsequent measurements of time are taken, which are used to 
calculate acceleration $\tilde {a}_\bot $ in the transverse direction, the 
latter measurements for time will be on clocks that are further out along 
the \textbf{k} direction than earlier clocks. In fact, by substituting (\ref{eq29}) 
into (\ref{eq32}), one sees the purely kinematic dependence of $\tilde {a}_\bot $ on 
$\tilde {a}_k $. And hence, the settings on the clocks in a non-E 
synchronization will modify what one would otherwise expect the measurement 
of $\tilde {a}_\bot $ to be. Certainly, we would expect this modification to 
be dependent on the magnitude of the R synchronization $k$ and acceleration 
$\tilde {a}_k $ in the \textbf{k} direction, as is found on the RH side of 
(\ref{eq32}). Further, $\tilde {a}_\bot $ is measured with meter sticks as well as 
clocks, and the more meter sticks the object passes in the transverse 
direction for a given change in the readings on the clocks the object 
passes, the greater the value of $\tilde {a}_\bot $. Thus, we would also 
expect $\tilde {a}_\bot $ to depend on velocity $\tilde {v}_\bot $, as it 
indeed does in (\ref{eq32})\cite{One:1}.

\subsection{Clocks, Acceleration, and Pseudo Forces}
\label{subsec:clocks}
We have seen there are no real pseudo forces, as the term is usually used, 
in the R system equations of motion. Though one might be tempted to conclude 
that the extra terms/factors in (\ref{eq25}) represent such pseudo forces, we see 
that a correction in the equation of motion arises to account for the 
different numbers measured for acceleration with different clocks. As the 
particle moves in the direction of non-Einstein synchronization, the 
different readings on the clocks along that direction change our numerical 
value for acceleration. The correction is found in the extra terms and 
factors of (\ref{eq25}), and thus, it seems more appropriate to call the function on 
the RH of (\ref{eq25}) a ``synchronization function'', rather than a ``force 
function''. There are no ``true'' pseudo forces, as in rotation, but only a 
clock correction, or ``pseudo'' pseudo force.

\subsection{Observation from the R System}
If we merely change the settings on the clocks -- similar to merely changing 
the settings (numbers, or labels) on the spatial coordinate grid - observers 
of any process (single particle motion, collision of particles, etc.), would 
see no intrinsic \textit{visual} difference in behavior. Coordinate values would change, 
but not observed behavior. No strange pseudo forces would seem to push 
objects in unexpected directions. Colliding billiard balls would look no 
different to our eyes (unlike in rotation where they would.) 

Further, the first order length changes calculated\cite{Ref:2} for the R 
system would also not manifest visually. This is because Lorentz contraction 
and other length changes, such as that considered here, are calculated in 
relativity by assuming the 3D endpoints of a given rod exist at the same 
moment in time\cite{straightforward:1} as seen by a given observer.

Consider a rod moving to the right as seen by both E and R observers. At 
$t=\tilde {t}=0$ the LH end of the rod is at $x=\tilde {x}=0$ For the E 
observer, the RH end is at $x=L$ for $t$ = 0. But the RH end according to the R 
observer (though she sees the same thing visually as the E observer) when it 
is at $\tilde {x}=L$ is not at $\tilde {t}=0$, but at some earlier time, 
according to the R clock at $\tilde {x}=L$. A short time later the RH end 
will have moved further out along the $x$ axis, and then the R clock where the 
RH end is located would read $\tilde {t}=0$. Taking the endpoints of the rod 
as simultaneous (existing at the same time) means the distance between the 
LH and RH ends of the rod according to the R system, at time $\tilde {t}=0$, 
is greater than $L$. But this difference is simply an artifact of arbitrarily 
setting the R clocks differently than the E clocks. Physically, an R 
observer actually perceives no visual difference in rod length from the E 
observer.

If one calculated the times of arrival of light rays from rod endpoints to 
an E and an R observer co-existent at the same 4D location, the calculations 
in the two systems would be different, but the conclusions as to what each 
would see visually would be the same.

And so it goes with all other ostensible physical differences between the R 
and E systems. They do not manifest as any difference to physical witnesses 
(unlike rotation, for example), but only in the equations to describe 
phenomena those witnesses would employ, based on their disparate choices for 
clock settings.

\section{Inertial or Non-inertial Frame?}
\label{sec:inertial}
Ohanian contends that, due to the arising of ``pseudo forces'', R 
synchronization coordinates constitute non-inertial frames. He does not 
discredit conventionality of synchronization completely, but posits that R 
systems cannot therefore be valid representations of inertial frames. 
Indeed, Ohanian says, ``.. the R frame is a possible reference frame for 
describing physics. [But because] Newton's laws of dynamics are not valid in 
their standard form {\ldots} the R frame \textit{is not an inertial frame..}''\cite{Ref:3}

Yet, one could argue that the most fundamental definition of a ``frame'' is 
a set of continuous 3D points, each of which keeps a constant spatial 
distance from every other such point. This is certainly true for the R 
coordinates. Further, ${\rm {\bf x}}={\rm {\bf \tilde 
{x}}}=\,\mbox{constant}$ for all points for all time, be it $t$ or $\tilde {t}$ 
time. Hence, there is never any motion between the two frames, and they must 
therefore be the same frame. They have different \textit{coordinates}, specifically the time 
coordinates, but the \textit{frame} is the same. This is analogous to a purely spatial 
coordinate change (such as $x\to x+3)$ for which the underlying frame stays 
the same. Thus, if the E system represents an inertial frame, so must the R 
system. 

Additionally, we must draw a distinction between \textbf{F} = m\textbf{a} 
acting on a particle seen in an inertial frame, and the force felt by an 
observer fixed to a frame that is non-inertial. Although the form of the 
second law describing particle motion changes in going from E to R, the 
force felt by any observer fixed to a given spatial coordinate point stays 
the same, i.e., zero. If one feels no force on one's own body, one is in an 
inertial frame.

Though, from this logic, the R system and the E system appear to be 
\textit{different coordinate} systems, yet constitute the \textit{same frame}, one's conclusion in this regard seems to 
depend on one's choice of definition for inertial frame.

Traditionally, physics employs several heretofore seemingly equivalent ways 
to define an inertial frame. These are listed in Table 1. The third method 
in the list is not usually found in texts, though I submit most physicists 
would agree that it is a valid means. Although the last column is for a 
rotating frame, any non-inertial (accelerating or gravitational) frame would 
do.

\begin{longtable}[htbp]
{|p{13pt}|p{170pt}|l|l|l|l|l|p{9pt}|}
\caption{. Inertial Frame Definition Comparison} \\
\hline
\endhead
\hline
\endfoot
\multicolumn{2}{|p{184pt}|}{\begin{center}
\textbf{Inertial Frame Definition Method} \end{center} } & 
\textbf{E Synch Sys}& 
 & 
\textbf{R Synch Sys}& 
 & 
\textbf{Rotating Frame}& 
\begin{center}
  \end{center}  \\
\hline
\begin{center}
1 \end{center} & 
Fixed observers feel no body force?& 
Y& 
& 
Y& 
& 
N& 
 \\
\hline
\begin{center}
2 \end{center} & 
Geodesics look like straight lines?& 
Y& 
& 
Y& 
& 
N& 
 \\
\hline
\begin{center}
3 \end{center} & 
Fixed 3D points are permanently stationary relative to a known inertial frame?& 
Y& 
& 
Y& 
& 
N& 
 \\
\hline
\begin{center}
4 \end{center} & 
$\begin{array}{l}
 \quad \\ 
 m\frac{d^2x^i}{d\tau ^2}=F^i \\ 
 \quad \\ 
 \end{array}?^{\ast }$& 
Y& 
& 
Y& 
& 
N& 
 \\
\hline
\begin{center}
5 \end{center} & 
$\begin{array}{l}
 \quad \\ 
 m\frac{d^2x^i}{dt^2}=F^i \\ 
 \quad \\ 
 \end{array}$ to 1st order?$^{\ast }$& 
Y& 
& 
N& 
& 
N& 
 \\
\hline
\multicolumn{8}{|p{322pt}|}{* Assumes Cartesian spatial grid} 
\label{tab1}
\end{longtable}

Four of the five criteria directly imply the R system is inertial. This 
includes the fully relativistic equation of motion (fourth method), though 
not what that typically reduces to at first order, i.e., Newton's second law 
for inertial frames (fifth method). Thus, though these two methods have 
usually been considered more or less equivalent, when R synchronization 
systems are considered, they are not. Of the two, most would consider the 
fully relativistic (4$^{th}$ method) to be the more fundamental.

The difference arises in the determination of coordinate standard clock time 
$t$ from the proper time on the particle $\tau $. In the R synchronization this 
has first order dependence on the velocity component in the direction of the 
non-Einstein clock synchronization. Again, I would argue that this is merely 
a \textit{coordinate} difference between the R and E systems, and does not imply they are 
different \textit{frames.} Thus, by any measure, the R system constitutes an inertial frame.

\section{Philosophy, Semantics, and Synchronization}
\label{sec:philosophy}
Virtually everyone agrees that the ``.. in the R [system] physics is 
consistent and complete.''\cite{Ref:4} Calculations, based on theory, can 
be made that predict observed phenomena. Further, everyone admits that the R 
system is not as ``pretty'', or economical, as the E system. After that, the 
arguments seem to drift to the philosophical, even semantic, rather than the 
scientific.

For example, if prior well-used, successful, and heretofore interchangeable 
definitions of ``inertial'' seem in conflict with one another, contending 
that one is more correct than another seems to miss the point. The old 
definitions must be re-thought and refined, in the context of the new 
knowledge that has arisen.

And whether we consider violation of a long-held sacrosanct definition, 
principle, or philosophical position to be enough to ``defeat'' the 
conventionalist thesis seems to be largely a matter of personal 
predilection. For example, Poincare' invariance has held prominence for 
decades as a seeming inviolate bedrock of natural law. When it, or its 
Newtonian sibling \textbf{F}=$m$\textbf{a}, are controverted by a new theory, 
one can feel a certain historical justification in rejecting such a theory 
on principle. However, there is no \textit{a piori} reason, no reason other than past 
experience with standard synchronization, upon which to base such a 
judgment. As it does not appear that the conventionalists have ever made any 
claim other than that their thesis is consistent, internally and with 
experiment, if we are to invalidate that thesis, then it seems that claim, 
and that claim alone, is where we should start.

\section{Conclusion}
\label{sec:conclusion}
With regard to non-standard Reichenbach synchronization gauges, dynamics is 
no different from kinematics. Kinematically, we know we are in a 
non-Einstein synched system because the measured one-way speed of light is 
anisotropic. Dynamically, we know because \textbf{F}=m\textbf{a} is not 
isotropic (or more precisely, the first order relativistic generalization of 
Newton's second law where time is local standard clock time, not proper time 
on the particle.)

Ohanian's discrediting of non-standard synchronization is right if we demand 
that nature's symmetry extend beyond the fully covariant tensor form (using 
proper time) of physical laws to that of our humanly chosen systems of 
clocks and rods. Then we must have Einstein synchronization. But that demand 
seems decidedly artificial (though more esthetic and simple, to be sure.) 
Using tensor notation or generalized coordinates, no new pseudo forces 
arise. Using non-standard synchronized clocks, the coordinate equations of 
motion change form, but that must be expected. In any case, the R system, as 
judged from almost any perspective, remains an inertial frame. And in either 
the E or R system, provided we use the same clock synchronization for both 
analysis and test, we will get theoretical predictions that match 
experiment. And in the end, that is all we can really ask of physics. 

I love symmetry, beauty, and simplicity in my physics, and the 
conventionalist view of synchronization, lacking, in my opinion, those 
qualities, is not something I am particularly enamored of. I would, in fact, 
be quite pleased if someone could find a way to do away with it, 
emphatically and finally. Thus, I initially welcomed, whole-heartedly, the 
seeming refutation of the conventionalist's thesis by Professor Ohanian, for 
whom I have long held considerable admiration. However, upon further 
reflection, I reluctantly concluded that, once again, an attempt to 
invalidate the theory of gauge synchronization seems to have come up short.


\begin{thebibliography}{6}
\bibitem{straightforward:1} It is straightforward to show, in Lorentzian relativity theory, that if the 3D endpoints of a rod are simultaneous (exist at the same time) as seen in two different systems, then there is no difference in length (no Lorentz contraction) for the rod between the two systems. Lorentz contraction is intimately related to differences in simultaneity (and thus to synchronization).
\end{thebibliography}
\end{document}